\documentclass[twocolumn]{aastex62}

\usepackage{amsmath}
\usepackage{color}
\graphicspath{{./}{figures/}}
\usepackage{natbib}
\usepackage{enumitem}
\usepackage{float}
\usepackage{hyperref}

\submitjournal{ApJ}
\usepackage{url}

\shorttitle{The Dragonfly Wide Field Survey I}
\shortauthors{Danieli et al.}

\begin{document}

\title{The Dragonfly Wide Field Survey. I. Telescope, Survey Design and Data Characterization}

\correspondingauthor{Shany Danieli}
\email{shany.danieli@yale.edu, shanyi1@gmail.com}

\author[0000-0002-1841-2252]{Shany Danieli}
\affil{Department of Physics, Yale University, New Haven, CT 06520, USA \\}
\affil{Yale Center for Astronomy and Astrophysics, Yale University, New Haven, CT 06511, USA \\}
\affil{Department of Astronomy, Yale University, New Haven, CT 06511, USA \\}

\author[0000-0002-2406-7344]{Deborah Lokhorst}
\affiliation{Department of Astronomy \& Astrophysics, University of Toronto, 50 St. George Street, Toronto, ON M5S 3H4, Canada\\}
\affiliation{Dunlap Institute for Astronomy and Astrophysics, University of Toronto, Toronto ON, M5S 3H4, Canada\\}

\author[0000-0001-5310-4186]{Jielai Zhang}
\affiliation{Centre for Astrophysics and Supercomputing, Swinburne University of Technology, P.O. Box 218, H29, Hawthorn, VIC 3122, Australia}

\author[0000-0001-9467-7298]{Allison Merritt}
\affiliation{Max-Planck-Institut f¨ur Astronomie, K¨unigstuhl 17, D-69117 Heidelberg, Germany}

\author[0000-0002-8282-9888]{Pieter van Dokkum}
\affiliation{Department of Astronomy, Yale University, New Haven, CT 06511, USA \\}

\author[0000-0002-4542-921X]{Roberto Abraham}
\affiliation{Department of Astronomy \& Astrophysics, University of Toronto, 50 St. George Street, Toronto, ON M5S 3H4, Canada\\}
\affiliation{Dunlap Institute for Astronomy and Astrophysics, University of Toronto, Toronto ON, M5S 3H4, Canada\\}

\author[0000-0002-1590-8551]{Charlie Conroy}
\affiliation{Harvard-Smithsonian Center for Astrophysics, 60 Garden Street, Cambridge, MA 02138, USA\\}

\author[0000-0002-8931-4684]{Colleen Gilhuly}
\affiliation{Department of Astronomy \& Astrophysics, University of Toronto, 50 St. George Street, Toronto, ON M5S 3H4, Canada\\}

\author[0000-0003-4970-2874]{Johnny Greco}
\affiliation{Center for Cosmology and Astroparticle Physics (CCAPP), The Ohio State University, Columbus, OH 43210, USA}

\author{Steven Janssens}
\affiliation{Department of Astronomy \& Astrophysics, University of Toronto, 50 St. George Street, Toronto, ON M5S 3H4, Canada\\}

\author[0000-0001-9592-4190]{Jiaxuan Li}
\affiliation{Department of Astronomy, Peking University, 5 Yiheyuan Road, Haidian District, Beijing 100871, China\\}

\author{Qing Liu}
\affiliation{Department of Astronomy \& Astrophysics, University of Toronto, 50 St. George Street, Toronto, ON M5S 3H4, Canada\\}

\author[0000-0002-5609-5243]{Tim B. Miller}
\affiliation{Department of Astronomy, Yale University, New Haven, CT 06511, USA \\}

\author[0000-0002-8530-9765]{Lamiya Mowla}
\affiliation{Department of Astronomy, Yale University, New Haven, CT 06511, USA \\}

\begin{abstract}

We present a description of the Dragonfly Wide Field Survey (DWFS), a deep photometric survey of a wide area of sky. The DWFS covers 330 $\mathrm{deg}^2$ in the equatorial GAMA fields and the Stripe 82 fields in the SDSS $g$ and $r$ bands. It is carried out with the 48-lens Dragonfly Telephoto Array, a telescope that is optimized for the detection of low surface brightness emission. The main goal of the survey is to study the dwarf galaxy population beyond the Local Group. In this paper, we describe the survey design and show early results. We reach $1\sigma$ depths of $\mu_g\approx 31$\,mag\,arcsec$^{-2}$ on arcminute scales, and show that Milky Way satellites such as Sextans, Bootes, and Ursa Major should be detectable out to $D\gtrsim 10$\,Mpc. We also provide an overview of the elements and operation of the 48-lens Dragonfly telescope and a detailed description of its data reduction pipeline. The pipeline is fully automated, with individual frames subjected to a rigorous series of quality tests. The sky subtraction is performed in two stages, ensuring that emission features with spatial scales up to $\sim 0\fdg9 \times 0\fdg6$  are preserved. The DWFS provides unparalleled sensitivity to low surface brightness features on arcminute scales.

\end{abstract}


\section{Introduction} \label{sec:intro}

Wide-angle imaging surveys, carried out in the past decade, have been one of the key drivers of progress in our understanding of extragalactic astronomy and cosmology. Notable examples include the Sloan Digital Sky Survey (SDSS; \citealt{2000AJ....120.1579Y}), the Galaxy And Mass Assembly surveys (GAMA; \citealt{2011MNRAS.413..971D}), the Kilo-Degree Survey (KiDS; \citealt{2019A&A...625A...2K}), the Dark Energy Survey (DES; \citealt{2018ApJS..239...18A}) and the Hyper Suprime-Cam Subaru Strategic Program (HSC-SSP; \citealt{2018PASJ...70S...4A}). This only makes an incomplete list of the extensive, high-quality data sets responsible for many important scientific breakthroughs in recent years. Taken together, they enable a multiwavelength characterization of large populations of astronomical sources, across cosmic time. These wide-field surveys exhibit a wide range in terms of instruments, filters, areas, and depth, as each is optimized for different science objectives.

Despite this progress, {\em surface brightness} remains a relatively unexplored parameter in imaging surveys. Surveys with photographic plates had demonstrated that a remarkable level of faint, extended structure and phenomena exist at surface brightness levels of $26-28 \ \mathrm{mag \ arcsec}^{-2}$ (e.g. \citealt{1971ApJ...163..195A}; \citealt{1971ApJ...169L...3W}; \citealt{1983ApJ...274..534M}; \citealt{1988ApJ...330..634I}). However, wide-field CCD cameras with mosaicked detectors are generally optimized for point-source depth and completeness rather than low surface brightness sensitivity. As such, modern surveys are not necessarily well suited for imaging extended, low surface brightness objects, such as the stellar halos of galaxies, low surface brightness galaxies, dwarf galaxies in the Local Volume out to $D\sim 15$\,Mpc, intragroup and intracluster light, and Galactic cirrus. A related issue is that the backgrounds are often difficult to measure on scales larger than an individual CCD. Having information on the large-scale background is important, as accurate measurements of the spatial extent and total luminosities of low surface brightness objects require accurate sky subtraction. 

This situation has recently changed owing to two separate advances. The first is that, thanks to advances in background subtraction and data analysis techniques, it has become possible to improve the preservation of low surface brightness emission in wide-field imaging data (\citealt{2010AJ....140..962M}; \citealt{2015MNRAS.446..120D}; \citealt{2015ApJ...807L...2K}; \citealt{2015ApJ...800L...3W}; \citealt{2016MNRAS.456.1359F}; \citealt{2016ApJ...823..123T}; \citealt{2017ApJ...834...16M}; \citealt{2018ApJ...857..104G}). Surface brightness levels as low as  $29-30 \ \mathrm{mag \ arcsec}^{-2}$ can now be reached in areas away from bright stars and chip edges, and faint galaxies are now routinely selected from general-purpose wide-field surveys (see \citealt{2018ApJ...857..104G}).

The second is the development of low surface brightness optimized telescopes, such as the Dragonfly Telephoto Array (`Dragonfly' for short). Dragonfly is a telescope that was designed to reach surface brightness levels below the $\sim 28-29 \ \mathrm{mag \ arcsec}^{-2}$ threshold that can routinely be reached with general-purpose telescopes  (\citealt{2014PASP..126...55A}). From 2014 to 2016 the telescope comprised eight excellently baffled telephoto lenses, mounted jointly and co-aligned to image the same area on the sky. Optically it operated as a $0.4 \ \mathrm{m}$ aperture $f/1.0$ refractor. Early Dragonfly results include the characterization of the outskirts of eight nearby Milky Way–like galaxies (\citealt{2016ApJ...830...62M}; \citealt{2018ApJ...855...78Z}), the identification and characterization of dwarf galaxies in nearby groups (\citealt{2014ApJ...787L..37M}; \citealt{2017ApJ...837..136D}; \citealt{2018ApJ...868...96C}), and the discovery of a large population of ultra-diffuse galaxies (UDGs) in the Coma Cluster (\citealt{2015ApJ...798L..45V}). Follow-up observations have contributed to the idea that low-mass galaxies show a wide variety in their dark matter content and stellar populations (\citealt{2018Natur.555..629V}; \citealt{2019ApJ...874L..12D}; \citealt{2019ApJ...874L...5V}; \citealt{2019ApJ...880...91V}; \citealt{2019arXiv191007529D}).

Despite its wide field of view of $\sim 6$\,deg$^2$, the eight-lens incarnation of Dragonfly was not an efficient tool for covering very wide areas, as long integration times were required to reach interesting depths (typically 20--30 hr per pointing). In 2017 the telescope was upgraded from 8 to 48 lenses,  making it optically equivalent to a 1\,m aperture, $f/0.4$ refractor with the same wide field of view as before ($2 \fdg 6 \times 1 \fdg 9$). This upgrade makes it possible to execute many-target or wide-area surveys efficiently. Among other programs, we are conducting the Dragonfly Edge-on Galaxy Survey (\citealt{2019arXiv191005358G}), which aims to characterize the stellar halos of galaxies, and the Dragonfly Wide Field Survey (DWFS), discussed in this paper.

The DWFS is intended to provide a unique data set for exploring the low surface brightness universe and in particular to probe the low-mass galaxy population beyond the Local Group, in groups and isolated environments. We have used the upgraded 48-lens Dragonfly to observe nearly $330 \ \mathrm{deg}^2$ to a depth equivalent to our earlier surveys (\citealt{2016ApJ...830...62M}), overlapping other multiwavelength data, covering regions in the Stripe 82 and GAMA fields.

We describe the 48-lens setup of Dragonfly in \S 2, followed by a description of its data reduction pipeline. We then describe the survey design and the observations in \S 4. In \S 5 we present preliminary results, including examples of low surface brightness sources, detected in the survey. We conclude with a summary and an outlook in \S 6.

\section{The 48-lens Dragonfly} \label{sec:dragonfly}
\subsection{Overview}

Dragonfly has evolved considerably from its initial configuration described in Abraham \& van Dokkum (\citeyear{2014PASP..126...55A}), so in this section we take the opportunity to update the reader with a description of the current configuration of the array. 

The Dragonfly concept is based on two distinct aspects. The first is the use of commercial high-end Canon telephoto lenses (400 mm $f/2.8$ Canon IS II). The lack of an obstruction in the light path and the use of all-refractive surfaces lead to a point spread function (PSF) that does not have complex high-frequency structure and has well-controlled wings. In principle, lenses scatter 10 times less light than mirrors at wide angles (\citealt{nelson}), though in practice ghosts and internal reflections from multiple lenses can lead to limitations in low surface brightness observations. These particular lenses have superb nanostructure-based anti-reflection coating and are very well baffled, which minimizes ghosting and flare caused by internal reflections (see \citealt{2014PASP..126...55A}). The second is the notion of co-aligning many of these lenses to simultaneously image the same position on the sky. In this manner, the effective aperture of the array, $D_{\mathrm{eff}}$, is that of an individual lens, $d_{\mathrm{individual}}$, increased proportionally to the square root of $n$, the number of lenses ($D_{\mathrm{eff}} = \sqrt{n} \cdot d_{\mathrm{individual}}$). The effective focal ratio is decreased by the same factor. Since the rate of detected photons per pixel, $\Phi$, varies inversely with the square of the focal ratio, arrays built out of a large number of lenses result in very efficient (optically fast) survey telescopes.

\subsection{Components}

\begin{figure*}[t!]
{\centering
  \includegraphics[width=0.99\textwidth]{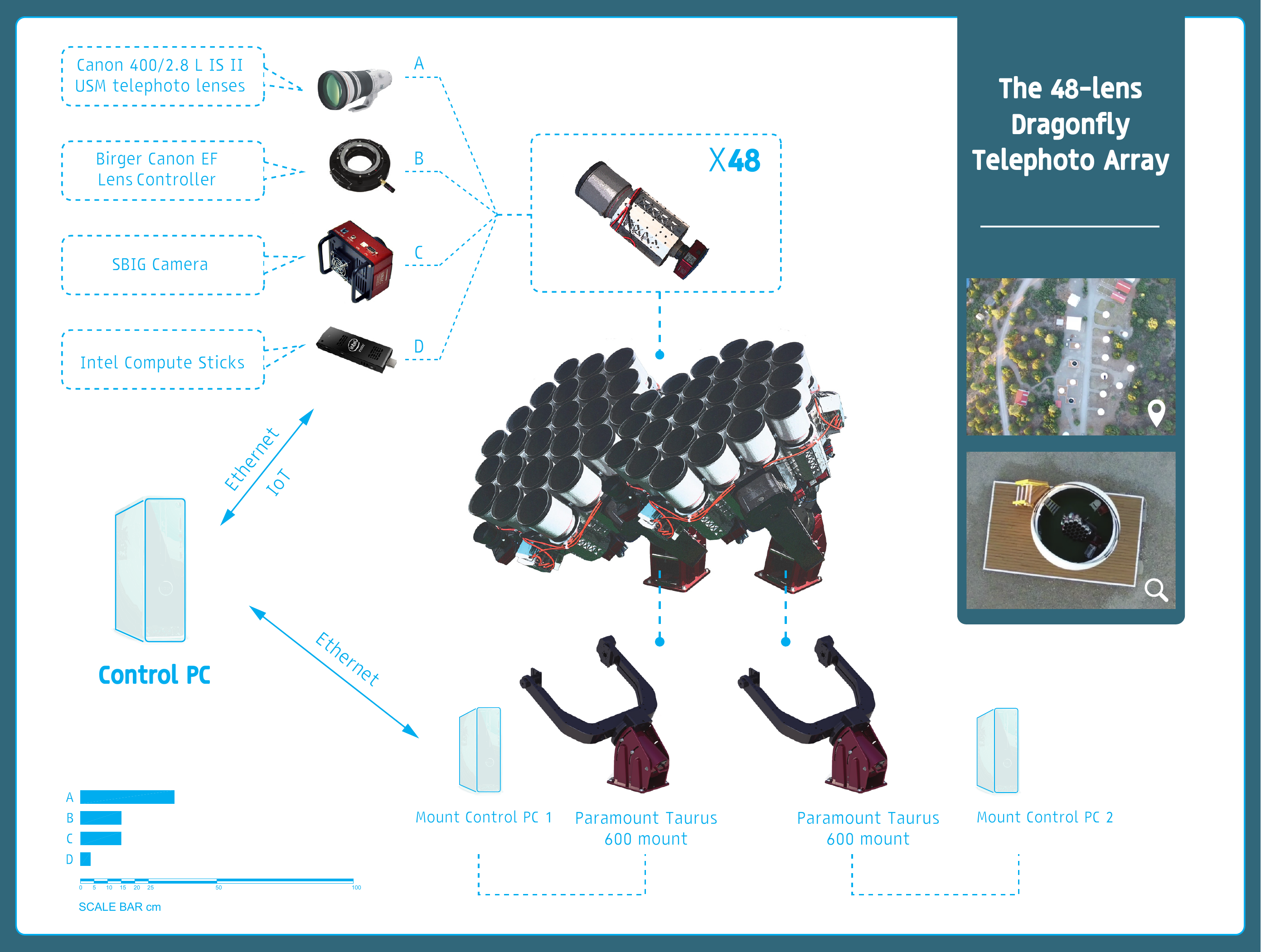}
  \caption{Schematic view of the 48-lens setup of the Dragonfly Telephoto Array that started routine operations in 2017. Each mount holds 24 lenses and is controlled by a separate mount control PC. Each lens (A) is a part of a fully independent subsystem with its own focuser (B), camera (C), and compute stick (D) and is controlled by a single control PC via an IoT protocol.}
  \label{fig:dragonfly}
}
\end{figure*}

A schematic overview of the current 48-lens Dragonfly is shown in Figure \ref{fig:dragonfly}. The Canon telephoto lenses constitute the heart of Dragonfly. A key element in the good performance of the array is the fact that the PSF of each lens is well behaved, with wings that are suppressed by a factor of $\sim 8$ compared to PSFs of conventional mirror telescopes. An implication of the lack of fine structure and steeply declining wide-angle PSF is that it can be modeled well. This is critical for the removal of high surface brightness compact sources, optimizing for an easier detection of faint structures (\citealt{2019arXiv191012867V}). Further information and characterization of the Dragonfly PSF will be provided in a pair of future papers (A. Merritt et al.
2020, in preparation; Q. Liu et al. 2020, in preparation).

Each lens in the array operates as a self-contained unit with its own internal auto-focusing motors. These motors (and, optionally, the aperture of the lenses) are controlled by custom units, made by Birger Engineering, which also serve as the physical connection between the lenses and the cameras. These cameras are Santa Barbara Imaging Group (SBIG) thermo-electrically cooled CCD cameras feeding Kodak KAF-8300 sensors, in various incarnations (model ST, SFT, or STT). When in focus the images are undersampled, with a stellar FWHM of $\sim 1.5 \ \mathrm{pixels}$ (see Figure 4 in Abraham \& van Dokkum \citeyear{2014PASP..126...55A}) and a pixel scale of $2\farcs85 \ \mathrm{pixel}^{-1}$ (corresponding to a physical pixel size of 5.4 \ $\mu\mathrm{m}$). Each lens is  equipped with a filter, contained within a drop-in filter holder. Half (24) of the lenses are equipped with Sloan-$g$  and half with  Sloan-$r$. Dragonfly therefore always takes simultaneous data in these two filters.

Apart from scaling up the number of lenses, several system improvements were made in the upgrade from 8 to 48 lenses. The first is the mounting of the array. The 48-lens+focuser+camera subsystems are mounted on two separate Paramount Taurus 600 mounts, manufactured by Software Bisque Inc., each holding 24 units. A second substantial upgrade has been to improve the operational protocols that are used to communicate with the 48 units. In its current configuration, each lens-focuser-camera system is connected to an Intel Compute Stick (a miniature computer), which is physically attached to the camera unit and connected through a secure local network to a central master control computer. Each unit is controlled by a node.js web server with a RESTful API, so operational commands are sent as URI strings by the master control computer to each one of the subsystems. The Internet of Things (IoT) architecture allows independent operation of each one of the subsystems, which adds to the robustness of the entire array; an error in one of the 48 units does not affect the operation of the other 47.

\subsection{Nightly Operation}
As described in Abraham \& van Dokkum (\citeyear{2014PASP..126...55A}), Dragonfly is located at the New Mexico Skies observatory\footnote{https://www.nmskies.com}. The nightly operation of Dragonfly is fully automated. At the beginning of each night, a Python script is executed that tests all of the system's components: the mounts, compute sticks, focusers, and cameras. Then, a shell script is executed that specifies the targets for observation and the required setup (exposure time and dither pattern). This script is either manually created or generated automatically from a target database. For the DWFS the script was generated manually.\footnote{We are moving toward a fully autonomous telescope, where everything from target selection to data validation and reduction is done without user intervention.} 

During twilight, a series of flat-field frames is taken, accompanied by a set of dark frames with the same exposure times as the flat-field frames. The exposure times of the flat fields are automatically adjusted to account for the expected change in sky brightness. At $\sim 15\arcdeg$ twilight the system is brought into focus with each lens performing an independent focusing procedure. This procedure involves obtaining short exposures at a series of focus settings and fitting a parabola to the measured widths of the stars to find the minimum. The telescope then executes the first science observing sequence, typically consisting of nine exposures of 600\,s each, dithered in a quasi-random pattern in a $45\arcmin$ box. The system actively microadjusts the focus based on the ambient temperature and repeats the full focusing procedure if needed. Each observing sequence is followed by a dark frame exposure with the same integration time as was used for a single science exposure. We note that science exposures for this project are only taken during dark time.
During morning twilight, another series of flats is obtained. During the night, frames (lights, darks, and flats) are temporarily stored on the individual compute sticks, and at the end of the night the data are synced and transferred to Redundant Array of Inexpensive Disks (RAID) storage, located at the University of Toronto. 

\section{Data Reduction} \label{sec:reduction}
Characterizing extremely low surface brightness emission requires excellent control of systematics, and we have developed a custom pipeline for processing the images observed for the DWFS. The most important systematics are flat-fielding, sky modeling and subtraction, and understanding the wide-angle PSF. Here we describe the Dragonfly reduction pipeline software architecture, as well as the cloud-based processing system that was set up and used in order to deal with the large amount of data taken with Dragonfly \footnote{Aspects of the pipeline have been described previously in Jielai Zhang's PhD thesis (\url{https://jielaizhang.github.io/files/Zhang_Jielai_201811_PhD_Thesis_excludech4.pdf}), and we refer the interested reader to this document for further information.}. Specific aspects relevant only for the DWFS data are discussed in \S\,\ref{survey.sec}.

\subsection{Infrastructure: Data Storage, Management, and Bookkeeping}\label{sec:infrastructure}
As described in Section \ref{sec:dragonfly}, each one of the lens-focuser-camera units is equipped with an Intel Compute Stick. Data collected by each unit are stored on the compute stick, and some basic data post-processing tasks take place immediately after each frame is obtained. First, data quality parameters are calculated for all images, using the Source Extractor software (SExtractor; \citealt{1996A&AS..117..393B}). Relevant parameters include the number of objects, their median FWHM, and their median axis ratio. These measurements are compared to predefined thresholds that require a minimum number of detected objects and a range for the FWHM. If a large fraction of cameras do not pass these tests, the frame is likely not in focus and the system performs a focusing sequence. Next, a preliminary WCS solution is computed using {\tt Astrometry.net} (\citealt{2010AJ....139.1782L}). Finally, postage stamps and the footprint from a full observing sequence, taking into account dithering and the intended misalignment between the lenses, are distributed to an email account that monitors progress. This allows observers to check whether the system is performing to specifications during the night and provides an easy-to-parse log that can be referenced if oddities in the data are discovered later.

After each night the raw data are copied from the RAID storage at the University of Toronto to two locations: the Canadian Institute of Theoretical Astrophysics (CITA) and the VOSpace, cloud-based storage provided by the Canadian Advanced Network for Astronomical Research (CANFAR). The VOSpace data storage \footnote{\url{http://www.canfar.net/en/docs/storage/}} is used in the following data processing stages and in the reduction pipeline (see \S\,\ref{sec:pipeline}). Data management is done through a comprehensive database that is automatically maintained.

\subsection{CANFAR Cloud Computing and Batch Processing} \label{sec:canfar_batch}
Besides data storage and management, the CANFAR cloud service is also used for all of the survey data processing. Different parts of the processing pipeline are executed in different environments. Steps that require large volumes are run on persistent cloud instances with substantial storage capacity and processing performance. A total of 10 virtual machines, each with four cores, 32 GB RAM, and 3-5 TB storage are dedicated to the project. Smaller and more common tasks are executed efficiently in a batch processing mode using the CANFAR resources. In this mode, some of the reduction pipeline tasks are processed in parallel by launching thousands of ``skinny" virtual machines for performing specific light tasks in parallel. The combination of the two described processing setups speeds up the entire reduction procedure, enabling the pipeline to keep up with the data flow from the telescope. The typical number of exposures collected by the 48 units in one night, including science, dark, and flat frames, is $\sim 3500$ frames (in good observing conditions).

\subsection{The Dragonfly Reduction Pipeline} \label{sec:pipeline}
In this subsection, we describe the (fully automated) steps of the Dragonfly reduction pipeline. We summarize the main algorithms and scripts that are the backbones of the data reduction software package. More detailed descriptions will be presented in J. Zhang et al. (2020, in preparation). The pipeline takes raw frames (dark, flat, and light exposures) and returns calibrated, sky-subtracted, registered, astrometrically calibrated, co-added images in the $g$ and $r$ bands.

Compared to standard reduction procedures, the Dragonfly pipeline has two unusual aspects. The first stems from the fact that Dragonfly automatically collects data in all conditions, whenever the domes are open. Consequently, many raw images exhibit issues due to cloudy weather, out-of-focus cameras, pointing errors, excessive scattering in the upper atmosphere, etc. The reduction pipeline was built according to the philosophy that bad frames are bad for many different reasons but good frames are all alike. Each frame (dark, flat, and light) is examined by a series of algorithms that serve as ``gates," only letting a frame pass if it satisfies certain criteria. Frames that do not pass a gate are rejected by the pipeline and not considered for the rest of the reduction process. These gates ensure that only frames with excellent quality and photometry are included in the final calibrated science product. The second aspect is a two-stage sky subtraction procedure that is crucial for reducing systematic errors that might originate in the sky modeling process.

\begin{figure*}[t!]
{\centering
  \includegraphics[width=0.8\textwidth]{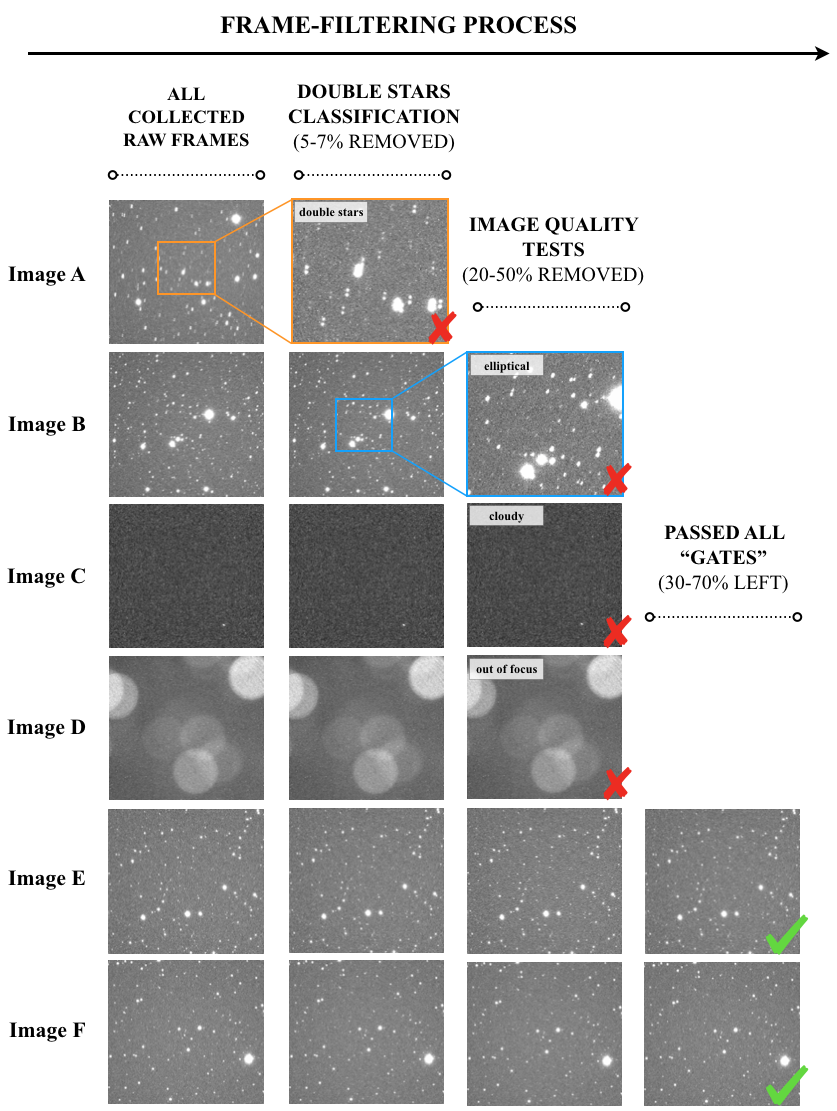}
  \caption{Gate-based image quality assessment in the Dragonfly reduction pipeline. Images that suffer from problems such as focusing errors, guiding errors, and clouds are identified and rejected as they progress through the pipeline.}
  \label{fig:pipeline}
}
\end{figure*}

\subsubsection{Classification of Calibration Frames and Construction of Master Darks and Master Flats}

Each of the 48 cameras has its own independent set of calibration frames. Darks are obtained every night with integration times that match those of flat-field and science frames.  All dark frames (up to $\sim 1500$ per night) are inspected and eventually classified as ``good" or ``bad" frames. Dark frames that are flagged as ``bad" are not used in the subsequent steps. Dark frames that are bad originate from problems such as hardware malfunctions that cause, for example, unplanned shutter opening and closing. Other issues can be caused by electromagnetic interference and by unstable temperature regulation of the camera. The last two can cause erroneous small-scale structure or large-scale gradients in the dark frames, respectively.

The rejection of frames that might experience such issues is done in the following way. For each dark frame, a 1D fifth-order polynomial function is fit to each column. Next, the resulting fit is median filtered in the $x$-direction. The image is then divided by this fitted, median-filtered model, and the resulting division frame is binned $10 \times 10$. The image is rejected if either (1) the ratio between the maximum and the minimum of the model is $> 1.1$, indicating a large-scale gradient, or (2) if the ratio of the maximum and the minimum value of the division image is $>1.015$, indicating small-scale structure, or (3) if the median or rms of the original frame is outside predefined bands of ``normal" expected values. Dark frames with the same exposure time that pass these dark inspection tests are average-combined into a master dark.

Dark-subtracted flat-field frames are also classified and combined into a master flat for each night and each camera. Individual flat frames are rejected if the Moon was above an altitude of $-3\arcdeg$ during the exposure, if the median pixel value of the flat is outside of a predefined range (i.e., too few or too many counts), or if linear streaks are present in the flat. Streaks are looked for by edge filtering the image using the Canny algorithm (in particular, we use the \texttt{feature} module of \texttt{scikit-image}\footnote{\url{https://scikit-image.org/}}). The detected edges are then run through a Hough transformation to look for lines. If the number of lines is larger than 80, the image is rejected. Flats are also rejected if a steep ramp is present in the data. First, individual flat frames are divided by a master flat created from flats that were not rejected thus far. Then, a linear model is fitted to a slice of the image after it was normalized and values larger than 1.1 were deleted. Individual flats are rejected if the best-fit model slope is steeper than $5\times 10^{-6}$. Flats are also rejected if they deviate too strongly from each other within a group (twilight or dawn). This is parameterized by the criterion that the standard deviation of all slopes in a group cannot be larger than $1\times 10^{-6}$. Lastly, if more than half of the single exposures in a group (twilight or dawn) have been rejected, the rest of the flats in the same group are also rejected by association. Flats that pass these tests are used to create a master flat by stacking at least seven dark-subtracted flat frames, for each subunit. If less than seven ``good" frames are available, no flat is created for that night.

\subsubsection{Rejecting Science Exposures with Doubly Imaged Sources}
Sometimes science exposures have two or more copies of every source in the image. This can happen when the telescope slews across the meridian,  causing a shift in the floating lens element that is used when image stabilization (IS) is engaged.\footnote{In our application IS is switched off on all the lenses, and partway through the data collection, the internal lens element was fixed in place with adhesive. Still, the internal floating lens element can occasionally shift.} An example can be seen in Figure \ref{fig:pipeline}. We deal with this (fairly rare) phenomenon by automatically inspecting each one of the individual science frames and rejecting those with ``double stars" upon identification. 

The algorithm for detecting these frames works as follows: First, SExtractor is run on the image in order to determine the positions of all sources, and a new image is created with a value of 1 at the central pixel position of all detected objects and 0 elsewhere. Then, an autocorrelation map of this image is created. Next, the image is flipped in $x$ and the autocorrelation of this mirrored image is obtained. The difference between the original autocorrelation image and the flipped one is $\sim 0$ if there are no double images, whereas there is a positive signal if there are double images. The central part ($200 \times 200$ pixels) of the difference image is examined, and the maximum value in this region is compared against a threshold that depends on the number of detected sources (this is done to prevent very crowded fields to be identified as having ``double stars," due to chance alignments). Frames that are identified as having double images are rejected and do not continue in the reduction process. Like the dark and flats classification, the classification of the double objects is processed in batch mode as described in Section \ref{sec:canfar_batch}. The Dragonfly Database includes information on whether the test was performed and what the result was for each raw frame.

\subsubsection{Dark Subtraction and Flat Fielding of Science Frames}
 All science frames that passed the first quality test are dark subtracted and divided by the matching master flat of the same subunit. In cases where no master flats were created for a specific night (due to the strict criteria on the quality of the flats), the closest temporally master flat is used. The dark frames that are used in the calibration of a science frame are generally taken on the same night. We verified that the CCD temperature shows seasonal variations but does not vary significantly during a single night. The calibrated science frames are then passed to the next steps of the reduction process.

\subsubsection{Science Image Quality Assessment and Rejection}\label{qualitytest}
This stage removes images that are out of focus, taken during adverse atmospheric conditions, or with the shutter or dome partially or fully closed. In this stage, SExtractor is run and several image properties are measured from the catalog of sources. The properties measured and the corresponding SExtractor keywords are as follows: the total number of objects detected in the image (NOBJ), the median FWHM (MEDFWHM), and the median ellipticity (MEDELLIP). Frames with a low number of objects ($\mathrm{NOBJ}<1000$) were probably taken in hazy conditions or when the telescope was out of focus, and these are rejected. Frames with  $\mathrm{MEDFWHM} > 5.5 \ \mathrm{pixels}$ due to bad focus and frames with $\mathrm{MEDELLIP}>0.3$ due to bad tracking are also rejected \footnote{A similar test, with more relaxed criteria, is done on the compute sticks straight after the data were taken, but no frames are rejected at that point. The test done on the sticks is done in order to inform the system whether it needs to refocus (see Section \ref{sec:infrastructure})}. Examples of image quality assessment and rejection are shown in Figure \ref{fig:pipeline}.  

\subsubsection{Rejection of Frames with Pointing Errors}\label{wrongpointing}

Very rarely, the telescope is not pointing at the target field owing to a failure in the control system. To remove these frames, \texttt{Astrometry.net} (\citealt{2010AJ....139.1782L}) is run and astrometric solutions are stored for all the light frames. A second-order polynomial is used for the distortion correction. Images with no WCS solution or images with a WCS solution with a central pointing that is more than $90 \arcmin$ from the target's coordinate are rejected. This step removes frames with actual pointing issues or with unrelated problems that cause the astrometry software to fail. Note that this is not the final astrometric solution; this is calculated at a later stage (see \S\,3.3.7).

\subsubsection{Sky Modeling and Model Subtraction - Stage I}
Accurately modeling the sky background emission is crucial for a reliable characterization of large-scale, low surface brightness phenomena such as stellar halos around massive galaxies, tidal features, and Galactic cirrus emission. We fit a model to pixels in each individual frame that are known to have no detected signal from astronomical sources in the final, combined frame. This requires a two-stage process, where the first stage serves to identify these ``empty'' pixels.

\begin{figure*}[t!]
{\centering
  \includegraphics[width=0.95\textwidth]{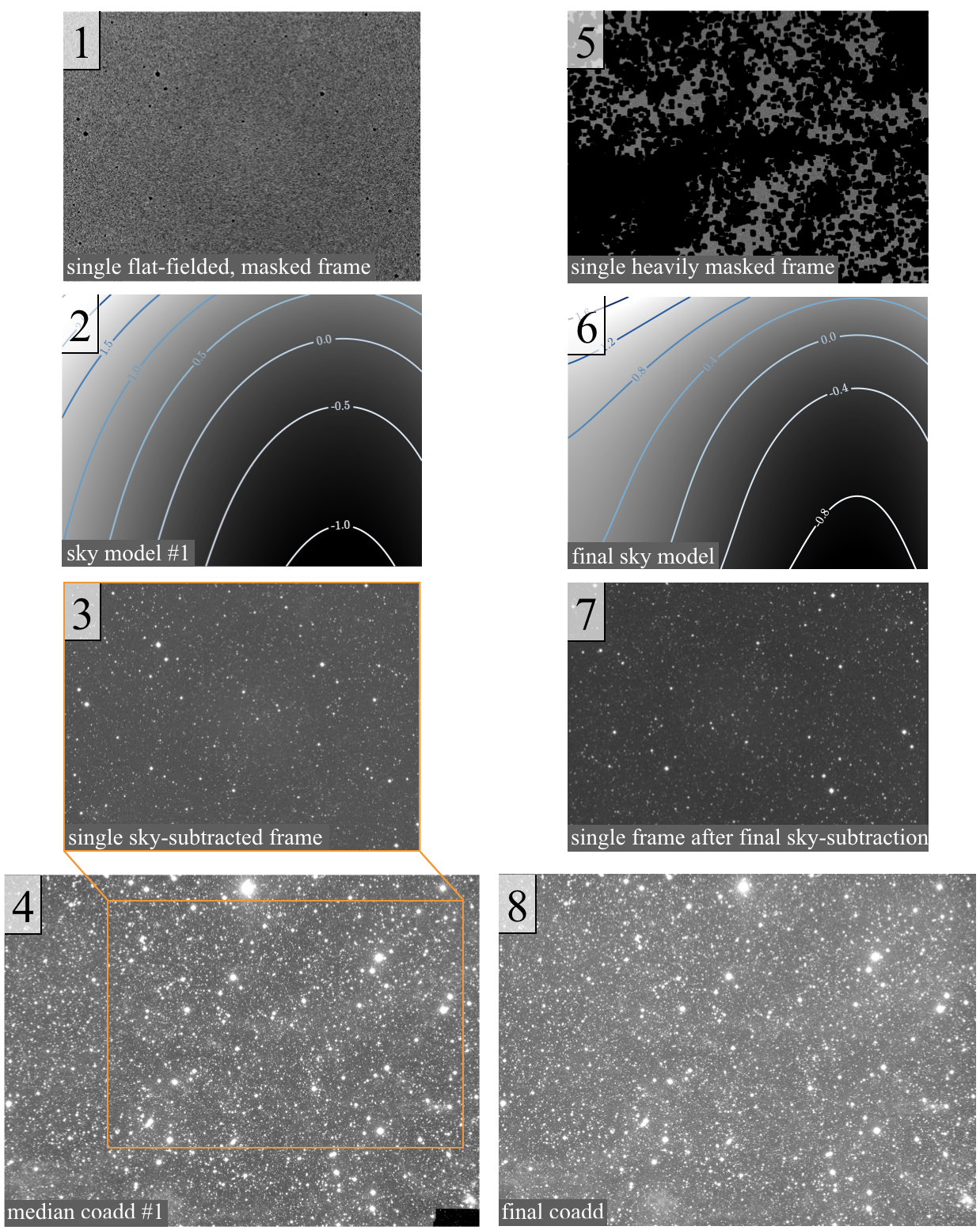}
  \caption{Illustration of the sky modeling and image combination. First, SExtractor is used to identify regions of the image where objects are and to generate a sky background with a $128\times 128$ pixel mesh (1).  This SExtractor sky is fitted with a third-order polynomial in $x$ and $y$ to generate the sky model (2 and 3). A first stacked image is generated by taking the median of all sky-subtracted frames (4). This median-combined image is used to create a very aggressive object mask, which includes all detected low surface brightness emission. This mask is projected back into each original frame (5), and the previous steps are repeated (6 and 7). The final combination is a mean, not a median, with masking of outliers (8).}
  \label{fig:sky-sub}
}
\end{figure*}

Here we describe the steps in the two-stage sky subtraction procedure while referring to Fig.\ \ref{fig:sky-sub} for a visual demonstration of the intermediate products. In the first stage, SExtractor is used to estimate the background of the image. This is done with a background mesh size of $128 \times 128$ (BACK\_SIZE$=128$). In the area of the background mesh, the mean and the standard deviation ($\sigma$) of the distribution are calculated while sigma-clipping in an iterative process until all the remaining pixel values are within $3\sigma$ from the mean. The final background map is a bicubic spline interpolation over the area of the mesh, after masking deviant pixels. A single masked frame used to estimate the background is shown in panel $1$ of Figure \ref{fig:sky-sub}. Next, a third-order polynomial is fitted to the background map, and the sky models are subtracted from the individual exposures. The sky model and a sky-subtracted individual exposure are shown in panels $2$ and $3$ of Figure \ref{fig:sky-sub}, respectively.

\subsubsection{Image Registration using SCAMP and SWarp}
Next, image registration of the sky-subtracted images is done using the software packages SCAMP (\citealt{2006ASPC..351..112B}) and SWarp (\citealt{2002ASPC..281..228B}). SExtractor is run, and a catalog of sources is created. First, SCAMP compares sources from the catalog to an online astrometric reference catalog, calculates a third-order astrometric solution, and stores it in a `.head' file for each individual image using WCS standards. SWarp then resamples the input images to a common grid using the stored astrometric solution. After resampling, the nonlinear distortions are smaller than $1''$ across the entire field of view, compared to $\sim 6''$ prior to correcting using SExtractor, SCAMP, and SWarp.

\subsubsection{Zero-point Calculation and Rejection of Nonphotometric Images}\label{stellarhalo}

The last stage of image rejection is done in order to ensure that no data taken under adverse atmospheric conditions enter the final co-add. This includes cloudy conditions and also scattering due to thin cirrus or other particles in the upper atmosphere. This scattering is particularly problematic, as it causes extended aureoles (``halos'') around all objects in the frame, severely compromising our ability to detect low surface brightness emission. While the impact of this systematic is mainly at wide angles, its presence is detectable from the changes it makes to the expected photometric zero-points. To account for both clouds and scattering, we require that the photometric zero-point of each frame falls within predefined limits.

The zero-point of each image is calculated by comparing the measured air-mass-corrected photometry of point source in the image to a reference photometric catalog. The default catalog is the AAVSO Photometric All-Sky Survey (APASS; \citealt{2016yCat.2336....0H}). These zero-points are then compared to the nominal ones, with each of the 48 cameras having its own reference zero-point. These reference zero-points are medians of large numbers of frames taken under known good conditions. Frames with air-mass-corrected zero-points that deviate from the nominal zero-point by more than a predefined value (typically 0.1\,mag) are rejected. This step also provides the photometric calibration of each frame.

\subsubsection{Image Combination I - Median Co-add}
The calibrated, sky-subtracted images are resampled to a common grid using SWarp and combined into two median images, in the $g$ and the $r$ bands. The two median images are summed to create a ``giant median co-add" (shown in panel $4$ of Figure \ref{fig:sky-sub}). This $g+r$ image serves as a model for the emission of astronomical sources in the field, down to the faintest levels probed by the full data sets.

\subsubsection{Sky Modeling and Model Subtraction - Stage II}
In the second stage of the sky modeling and subtraction, a similar procedure of estimating the background for each individual exposure is followed as in the first stage. SExtractor is used to estimate the background in a $128 \times 128$ mesh. However, in this stage, a weight map is given as an input for the masking stage. The weight map is created from the deep $g+r$ image described above, projected to the astrometric frame of each individual exposure. Sky pixels (to be fitted) have the value of 1 in the weight map where source pixels (to be masked) have the value 0. An example of individual masked exposure is shown in panel $5$ of Figure \ref{fig:sky-sub}. A background map is then created again for each individual frame and subtracted (panels $6$ and $7$ of Figure \ref{fig:sky-sub}, respectively).

This procedure was inspired by methods that are commonly used in the reduction of near-infrared imaging data (see, e.g., \citealt{2003AJ....125.1107L}), and it ensures that the background in the final combined frame is not oversubtracted near large and bright objects. The use of a third-order polynomial for individual frames means that we are not sensitive to structures with areas that exceed $\sim 0\fdg9 \times 0\fdg6 \sim 2000$\,arcmin$^2$.

\subsubsection{Final Image Combination}\label{combination}
After the second-stage sky subtraction, the images are resampled to a common grid using SWarp. Before stacking the images, a bad pixel map is generated in order to handle satellite trails and cosmic rays. In particular, satellite trails are the only ``bad" feature that occurs in all 48 lenses at once, and they are not reduced by the system redundancy. As such, they are much harder to remove than cosmic rays and other single-camera artifacts. The bad pixel map is generated as follows. First, the median image is subtracted from individual images to create a residual image. Both the median image and the individual images are flux scaled with a scale factor $S_{\mathrm{flux},i} = (\mathrm{flux \ scale})^{-1}$. Next, a noise model is created:

\begin{equation}
    \mathrm{noise}_{i} = \frac{1}{\mathrm{gain}}\sqrt{\mathrm{gain} \cdot \mathrm{median}(S_{\mathrm{flux},i} }) \cdot \frac{1}{S_{\mathrm{flux},i}}
\end{equation}

In the next step, the residual images are divided by the noise models. The resulting image is used to create a mask with pixels deviating more than $3\sigma$ and the mask is grown by 1 pixel in all directions to ensure masking of the faint edges of the satellite trails. Next, in order to unmask the centers of bright stars but keep the satellite trails and cosmic rays masked, the residual image is divided by a slightly smoothed ($3 \times 3$) median image to create a normalized residual image. In the normalized residual image, residuals from stars are expected to have low relative flux, while satellite trails and cosmic rays are expected to stand out. The normalized residual image is examined, and areas with values that are smaller than 1 in absolute value are unmasked. The final mask is saved as a bad pixel map for each individual image. Before turning to the final stacking, the fraction of rejected pixels per frame is calculated. Images with too few ($\mathrm{frac}<0.001$) or too many ($\mathrm{frac}>0.1$) rejected pixels are not included in the final combined image.

In order to optimize the signal-to-noise ratio in the combined images (by filter), a weighted average combination of individual exposures is created:
\begin{equation}
  I = \frac{\sum_{i=1}^{i=N}  w_i I_i}{\sum_{i=1}^{i=N} w_i} = \frac{\sum_{i=1}^{i=N}  \frac{I_i}{\mathrm{bg}_{_i} S_{\mathrm{flux},i}} } {\sum_{i=1}^{i=N} \left( {\rm bg}_i
  S_{\mathrm{flux},i} \right)^{-1}},
\end{equation}
with $\mathrm{bg}_i$ and $S_{\mathrm{flux,i}}$ being the sky brightness at the time of exposure and the applied zero-point flux scaling (which is inversely proportional to the image fluxes) in exposure $i$, respectively.

\subsection{Multi-resolution Filtering} \label{sec:mrf}

Dragonfly data generally have excellent low surface brightness sensitivity but suffer from crowding due to the FWHM\,$\approx 5\arcsec$ PSF. This crowding makes it difficult to distinguish low surface brightness objects from clumps of faint stars and galaxies. We have developed a method, multi-resolution filtering (MRF), to remove all compact emission sources from Dragonfly data.

The method is discussed in detail in van Dokkum et al.\ (\citeyear{2019arXiv191012867V}). Briefly, SExtractor is run on an independent image of higher spatial resolution, such as public imaging from CFHT, the VLT Survey Telescope, or Subaru. The segmentation map is converted to a mask and multiplied by the image to create a flux model of all detected sources. Low surface brightness objects and saturated stars are removed from this model. The model is then convolved with a kernel to match the Dragonfly PSF and subtracted. Bright stars are subtracted in a separate step, using the median of scaled images of these stars in the Dragonfly image. The final residual image contains only low surface brightness objects on the scale of the Dragonfly PSF, to a well-defined limit. The {\tt mrf} code, including examples, is available from {\tt github}.\footnote{\url{https://github.com/AstroJacobLi/mrf}}

\section{Survey Design, Observations, and Reduction}
\label{survey.sec}

In the following sections, we describe the rationale of the DWFS, the required surface brightness depth, and the selection of the target fields. The survey makes use of Dragonfly's ability to image the sky to very low surface brightness levels on large scales and, thanks to its built-in redundancy, its low incidence of artifacts and image defects. The DWFS is complementary to other wide-field studies that are optimized for point-source depth, spatial resolution, and/or multiwavelength coverage. Examples of such surveys that overlap with the DWFS footprint are the Kilo-Degree Survey (KiDS; \citealt{2019A&A...625A...2K}) and the Hyper Suprime-Cam SSP Survey (\citealt{2018PASJ...70S...4A}).

\subsection{Main Science Objective} \label{sec:objectives}

The DWFS was designed and initiated with the goal of better characterizing the faint end of the galaxy population outside of the Local Group. Faint galaxies are far more common than luminous ones and provide important constraints on structure formation and dark matter models. Perhaps the best-studied low-luminosity galaxies are the ``classical" dwarf spheroidals in the Local Group. These galaxies are dark matter dominated at all radii, and their density profiles have been used to constrain dark matter models (e.g. \citealt{2008ApJ...681L..13B}; \citealt{2009ApJ...704.1274W}; \citealt{2013MNRAS.429L..89A}). Galaxies with even lower luminosities ($L\lesssim 10^5  L_{\odot}$) are referred to as ultra-faint dwarfs (UFDs). These galaxies are extremely dark matter dominated, and an explanation for their faintness may be supernova feedback at very early times (\citealt{2018ApJ...863..123B} and references therein). Studying their chemical abundances and star formation histories is critical for understanding what processes cause low-mass galaxies to shut off their star formation and set a mass threshold for quenching of these galaxies (\citealt{2012ApJ...757...85G}; \citealt{2014ApJ...796...91B}; \citealt{2015ApJ...804..136W}).

As is well known, the luminosity function (and velocity function) of galaxies in the Local Group does not match simple expectations based on the halo mass function. At the lowest luminosities there is the ``missing satellite" problem, which may find its explanation in a combination of incompleteness and a low baryon content (\citealt{2008ApJ...688..277T}; \citealt{2014ApJ...795L..13H}; \citealt{2018MNRAS.473.2060J}; \citealt{2018PhRvL.121u1302K}; \citealt{2018MNRAS.479.2853N}). At higher luminosities, the counts are also below expectations (the ``too big to fail" problem in groups, \citealt{2011MNRAS.415L..40B}; and in the field, \citealt{2015A&A...574A.113P}). The ``too big to fail" problem in groups may result from stochastic variation between galaxy groups: early results from the SAGA survey show a large variety in the satellite populations of Milky Way-like galaxies (\citealt{2017ApJ...847....4G}).

Low-luminosity galaxies outside of the group environment may represent an entirely distinct population. It is likely that their formation and evolution are strongly environment dependent as will be reflected in their stellar kinematics, star formation histories, and chemical composition. Detecting the lowest-luminosity galaxies beyond the Local Group is difficult, not just because of their low integrated luminosities but also due to their low surface brightnesses of $\mu > 27 \ \mathrm{mag \ arcsec}^{-2}$. State-of-the-art deep, wide-angle conventional ground-based telescopes have successfully used individual giants stars to identify low-luminosity galaxies out to $4-5 \ \mathrm{Mpc}$ (e.g.,  Smercina et al.\ \citeyear{2018ApJ...863..152S}), but due to the $D^2$ fall-off of the brightness of point sources, it is difficult to extend these studies to larger distances.

As shown in Danieli et al.\ (\citeyear{2018ApJ...856...69D}), at distances $\gtrsim 5$\,Mpc it is likely easier to detect low-luminosity galaxies by their integrated low surface brightness glow than by the light of their individual stars. We have initiated the DWFS with the principal goal of constructing an optically selected sample of faint and ultra-faint field dwarf galaxies. Simply counting the number of such galaxies in isolation can place limits on the stellar mass--halo mass relation and on dark matter models (see, e.g., \citealt{2018ApJ...856...69D}).
An important aspect of studying such a sample is the ability to reliably measure their distances. Although the Dragonfly data themselves can be used to test specific models, distances are crucial for determining physical properties of individual galaxies. Greco et al. (\citeyear{2020arXiv200407273G}) provide an in-depth study of the prospect of using surface brightness fluctuations to determine distances to low surface brightness, low-mass galaxies. Other follow-up studies of their morphologies, colors, kinematics, and other properties will shed light on the masses and star formation histories of these galaxies and test whether there are thresholds for star formation at the lowest luminosities (\citealt{2015MNRAS.454.1798K}).

\subsection{Additional Science Goals}

The data products will be made publicly available, and in addition to our primary science goal, we expect that the DWFS will contribute to many other research areas. We provide an incomplete list of examples below.

\subsubsection{Tidal Tails and Streams}

As predicted by $\Lambda \mathrm{CDM}$ (\citealt{2005ApJ...635..931B}; \citealt{2008ApJ...689..936J}; \citealt{2010MNRAS.406..744C}), we expect to detect the signature of accretion events and galaxy mergers in the form of tidal distortions and tails, streams, and shells (e.g. \citealt{2013ApJ...765...28A}; \citealt{2018ApJ...866..103K}; \citealt{2019ApJ...883L..32V}). Such dynamical features serve as unique tracers for structure formation and carry information about the gravitation potential of dark matter halos on small scales (e.g. \citealt{1999ApJ...512L.109J}; \citealt{2010ApJ...712..260K}; \citealt{2019ApJ...880...38B}). Studies of stellar streams in the Milky Way have been used to provide constraints on the potential of the Galaxy and its accretion history (\citealt{2010ApJ...714..229L}; \citealt{2010ApJ...711...32N}; \citealt{2014MNRAS.445.3788G}; \citealt{2015MNRAS.449.1391B}; \citealt{2015ApJ...803...80K}; \citealt{2017ApJ...847...42D};  \citealt{2018ApJ...867..101B}). However, they also revealed, along with simulations, that most of these tidal features have surface brightness fainter than $29 \ \mathrm{mag \ arcsec}^{-2}$ and exhibit a rich variety of low surface brightness substructure (\citealt{2018MNRAS.480.1715P}). With this survey, we hope to provide data that will be able to increase the census of these dynamical events. 

\subsubsection{Low Surface Brightness Galaxies}
In the 1980s it was found that many galaxies were missed from traditional imaging surveys owing to their low surface brightness (e.g., \citealt{1976Natur.263..573D}). Many of these objects turned out to be low-density gas-rich spiral galaxies with faint, large disks  (e.g., \citealt{1993AJ....106..548V}; \citealt{2001ApJ...552L..23D}; \citealt{2013AJ....146...41S}). The most spectacular examples are Malin I (\citealt{1993ApJ...417..114S}) and other giant spirals (\citealt{1995AJ....109..558S}). These galaxies tend to live in fairly low-density environments and should be readily detectable in the survey.

\subsubsection{Ultra-diffuse Galaxies}

Not all large low surface brightness objects are spiral galaxies. Examples of extended spheroidal low surface brightness galaxies have also been known since the 1980s (\citealt{1984AJ.....89..919S}; \citealt{1997ARA&A..35..267I}) and were later found in small numbers in galaxy surveys (\citealt{1987AJ.....94.1126C}; \citealt{1988ApJ...330..634I}; \citealt{1997AJ....114..635D}; \citealt{2003ApJ...591..167C}). In 2015 it was found that they are surprisingly common, as the Dragonfly telescope identified them in large numbers in the Coma Cluster (\citealt{2015ApJ...798L..45V}). These UDGs have sizes of $R_{\rm e}>1.5$\,kpc and central surface brightness $\mu_g>24$\,mag\,arcsec$^{-2}$. The unexpected ubiquity of these galaxies, with thousands now known in Coma and other clusters (\citealt{2015ApJ...807L...2K}; \citealt{2017A&A...607A..79V}), led to extensive follow-up work. In terms of their luminosity distribution, cluster UDGs appear to be a continuous extension of the ``normal" galaxy population (\citealt{2019ApJ...875..155D}), but their dark matter content shows large variation (\citealt{2016ApJ...819L..20B}; \citealt{2019ApJ...874L..12D}; \citealt{2019ApJ...874L...5V}, \citeyear{2019ApJ...880...91V}). Furthermore, they appear to have unusual globular cluster populations (\citealt{2016ApJ...822L..31P}; \citealt{2017ApJ...844L..11V}; \citealt{2018MNRAS.475.4235A}; \citealt{2018ApJ...862...82L}).

Most of what we know about UDGs is based on galaxies in clusters and/or groups, as in those environments distances can be assigned with well-controlled probabilities (see, e.g., \citealt{2017A&A...607A..79V}), and this enables efficient follow-up studies. Many models for the formation of UDGs hypothesize an environmental dependence, such that isolated UDGs have a different formation channel or may not even exist (\citealt{2019MNRAS.485..382C}). The closest field UDGs could be missed in other surveys because of their large apparent size: the standard mesh size of SExtractor is $64\times 64$ pixels, corresponding to $16\arcsec\times 16\arcsec$ for a pixel size of $0\farcs 25$. As a result, galaxies with half-light radii of $>1.5$\,kpc could easily be removed in the sky subtraction stage at distances $D\lesssim 20$\,Mpc.

\subsubsection{Stellar Halos and Intragroup Light}
The faint structured stellar halos around central galaxies hold key information about past accretion events and perturbers throughout the history of massive galaxies. Previous studies that used stacked images have revealed a red color excess and excess light at large radii of such galaxies (\citealt{2010MNRAS.405.2697B}; \citealt{2011ApJ...731...89T}), a large scatter in stellar halo mass fractions around spiral galaxies (\citealt{2016ApJ...830...62M}), and a connection between the structures of massive galaxies and their dark matter halos (\citealt{2018MNRAS.475.3348H}; \citealt{2018MNRAS.480..521H}). The wide and deep nature of the survey should allow the study of stellar halos around massive galaxies in various mass ranges and across environments (see J. Li et al. 2020, in preparation).

Along similar lines, the extended halos of bright galaxies in galaxy groups and clusters, are themselves surrounded by a diffuse light component called intragroup and intracluster light (IGL and ICL, respectively). Studying the IGL and ICL is essential for a full characterization of the assembly history of groups and clusters as well as the characterization of their tidal fields, responsible for the stripping of stars from their host galaxies into the intragroup and intracluster mediums (\citealt{2005ApJ...631L..41M}; \citealt{2012MNRAS.425.2058B}; \citealt{2014MNRAS.437.3787C}; \citealt{2016IAUS..317...27M}).

\subsubsection{Accurate Photometry and Sizes of Bright Galaxies}

Photometrically derived mass and size estimates of galaxies rely heavily on performing accurate photometry in imaging surveys. Systematic errors in total fluxes can translate directly to biases in the inferred total masses and colors of galaxies, which in turn propagate into errors in derived mass functions and other derived relations (\citealt{2011MNRAS.418.1587T}; \citealt{2013MNRAS.435...87M}; \citealt{2014MNRAS.443..874B}; \citealt{2018PASJ...70S...6H}). 

Possibly the most fundamental contribution of the survey is that it provides accurate backgrounds over the survey area on scales $<0.^{\circ}6$. This allows point-source optimized surveys on other telescopes to calibrate their photometry and derive correction maps to their sky subtraction algorithms. The combination of high-resolution, deep data from Subaru, the VLT survey telescope, and other state-of-the-art facilities with the Dragonfly data should provide accurate total magnitudes and sizes of all objects in the overlapping areas.

\begin{figure*}[t]
{\centering
  \includegraphics[width=\textwidth]{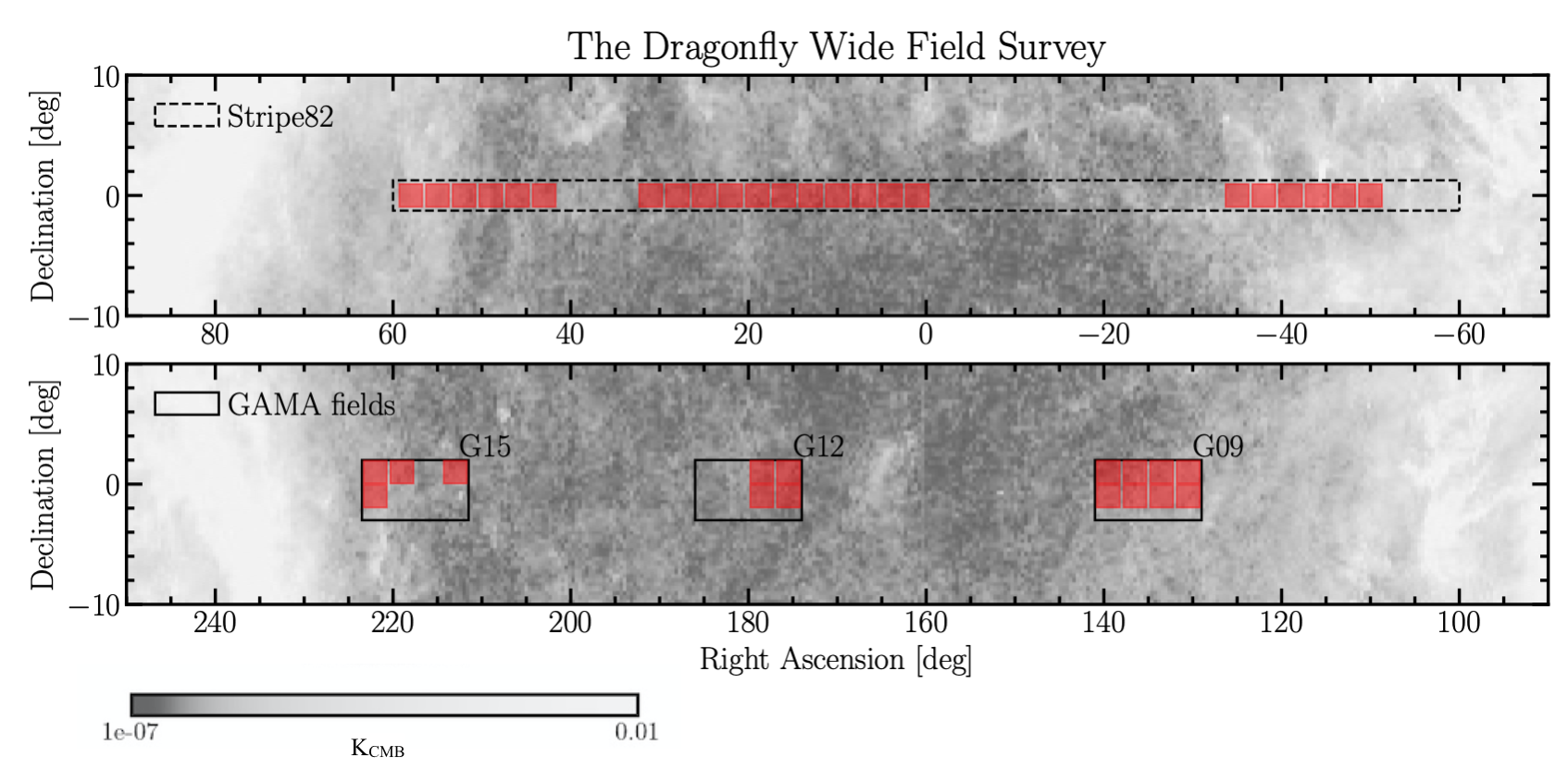}
  \caption{Dragonfly Wide Field Survey Y1 footprint on the sky in equatorial coordinates (red), covering areas of the GAMA fields (black solid line) and Stripe 82 (black dashed line). The footprint is overplotted on a  $353 \mathrm{GHz}$ dust map (\citealt{2011A&A...536A...1P}), with brighter areas indicating more dust.}
  \label{fig:footprint}
}
\end{figure*}

\subsection{Rationale of the Survey Depth and Covered Area} \label{sec:rationale}

The survey design was driven by the main science objective of characterizing the faint galaxy population.  In Danieli et al. (\citeyear{2018ApJ...856...69D}) we presented a model for calculating the predicted detection rates of dwarf galaxies using integrated light surveys, outside the Local Group.  As can be seen in Figure 4 of Danieli et al. (\citeyear{2018ApJ...856...69D}), the minimal detectable stellar mass out to $10 \ \mathrm{Mpc}$ strongly depends on the limiting surface brightness. The limiting stellar mass changes dramatically, ranging from stellar masses of $\sim \ 10^8 M_{\odot}$ for limiting surface brightness of $\sim \ 24 \ \mathrm{mag \ arcsec}^{-2}$ to stellar masses as low as $\sim \ 10^5 M_{\odot}$ for limiting surface brightness of $\sim \ 30 \ \mathrm{mag \ arcsec}^{-2}$ on scales of $10 \arcsec$.

Our goal is to reach a minimal mass of $\sim 10^5 M_{\odot}$ at 10 Mpc on $10 \arcsec$ scales with the assumptions we made in the Danieli et al. (\citeyear{2018ApJ...856...69D}) study. That requires a limiting surface brightness of $\mu_{\mathrm{eff,lim}}=29.5 \ \mathrm{mag \ arcsec^{-2}}$ in $V$ band. 
The number density of such objects in the volume between $3$ and $10 \ \mathrm{Mpc}$ is $0.05$ or $0.02 \ \mathrm{deg}^{-2}$, depending on the form of the stellar mass--halo mass relation. We therefore need to survey at least $350 \ \mathrm{deg}^2$ to distinguish between these possibilities: that will give either $\sim 17$ or $\sim 7$ objects in this mass range in the entire survey.

\subsection{Survey Fields and Observations} \label{sec:observations}

The locations of the DWFS fields were chosen to have considerable overlap with other surveys. As explained in \S\,\ref{sec:mrf}, we need high-resolution data with excellent point-source depth to isolate the low surface brightness emission. Furthermore, the availability of high-quality imaging at other wavelengths, as well as extensive spectroscopy, greatly aids in the interpretation of faint objects detected with Dragonfly. Finally, equatorial fields are beneficial for overlap with other surveys and accessibility from both hemispheres. Hence, the fields were chosen to overlap the footprint of the equatorial GAMA fields (\citealt{2011MNRAS.413..971D}) and the Stripe 82 field (\citealt{2009ApJS..182..543A}).

Observations for the DWFS were obtained between October 2017 and March 2019 with the 48-lens Dragonfly Telescope (with data for other programs being taken during this time frame as well). Over this period data were obtained in a wide range of conditions; as explained in \S\,\ref{sec:reduction} the pipeline automatically retains only good frames. The 48 lenses are offset from one another by $\approx 10 \%$ of the field of view, giving 48 independent sightlines. The data were typically taken in sequences of nine $600 \mathrm{s}$  exposures with large dithers of $\approx 45'$ between exposures. Each of the individual pointings covers $12 \ \mathrm{deg}^2$, after dithering, with reduced depth near the edges of the field. Figure \ref{fig:footprint} provides a view of the survey footprint. Also shown are the footprints of the equatorial GAMA fields and of Stripe 82.

\section{Preliminary Results}

\begin{figure*}[t!]
{\centering
  \includegraphics[width=\textwidth]{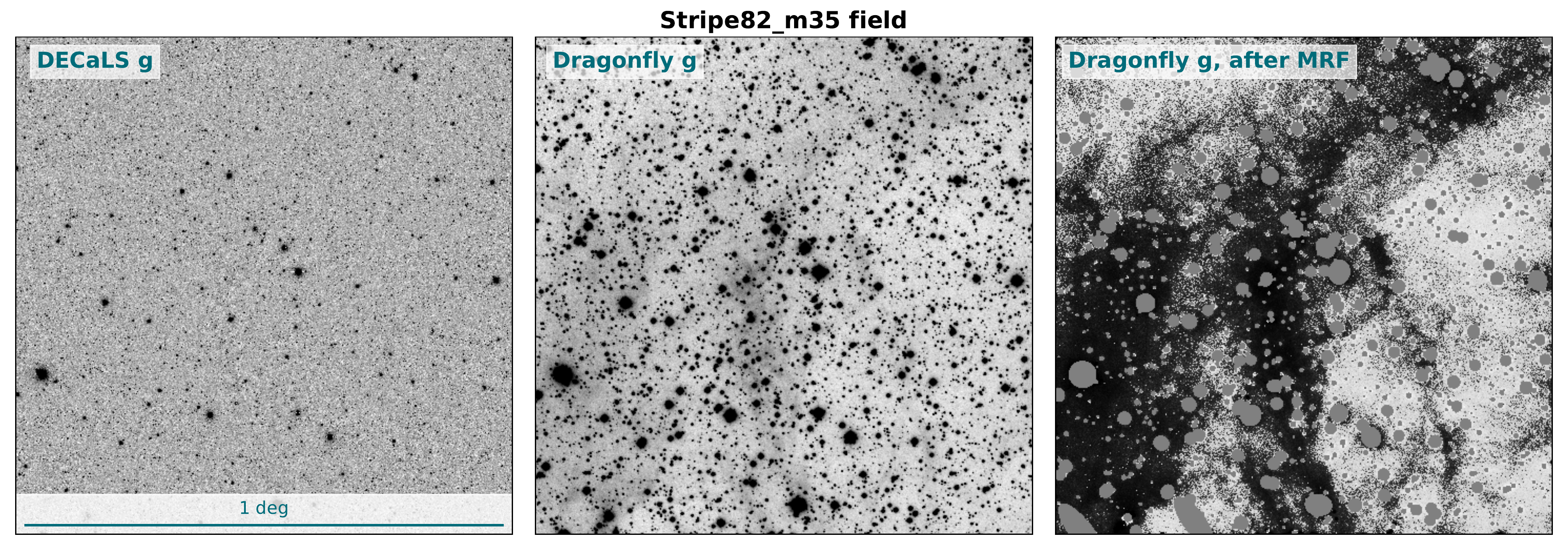}
  \caption{A $1 \times 1 \ \mathrm{deg}^2$ cutout of the survey field, Stripe82\_m35 (middle). The DECaLS data used in the MRF are shown in the left panel, and the Dragonfly data after applying the MRF are shown in the right panel. The degree-size structure is galactic cirrus, easily detected in the Dragonfly data.} 
  \label{fig:cirrus}
}
\end{figure*}

\begin{figure*}[t!]
{\centering
  \includegraphics[width=\textwidth]{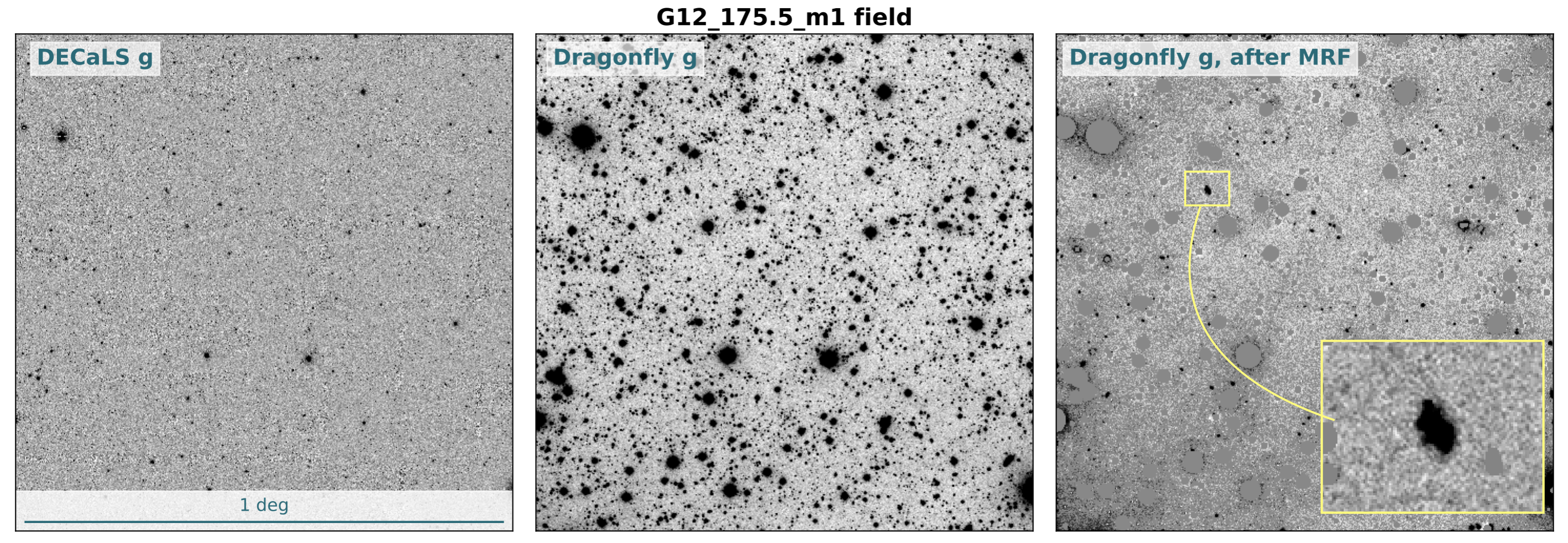}
  \caption{A $1 \times 1 \ \mathrm{deg}^2$ cutout of the survey field, G12\_175.1\_m1 (middle). The DECaLS data used in the MRF are shown in the left panel, and the Dragonfly data after applying the MRF are shown in the right panel. The inset shows a dwarf galaxy identified in the image.}
  \label{fig:g12}
}
\end{figure*}

\begin{figure*}[t!]
{\centering
  \includegraphics[width=\textwidth]{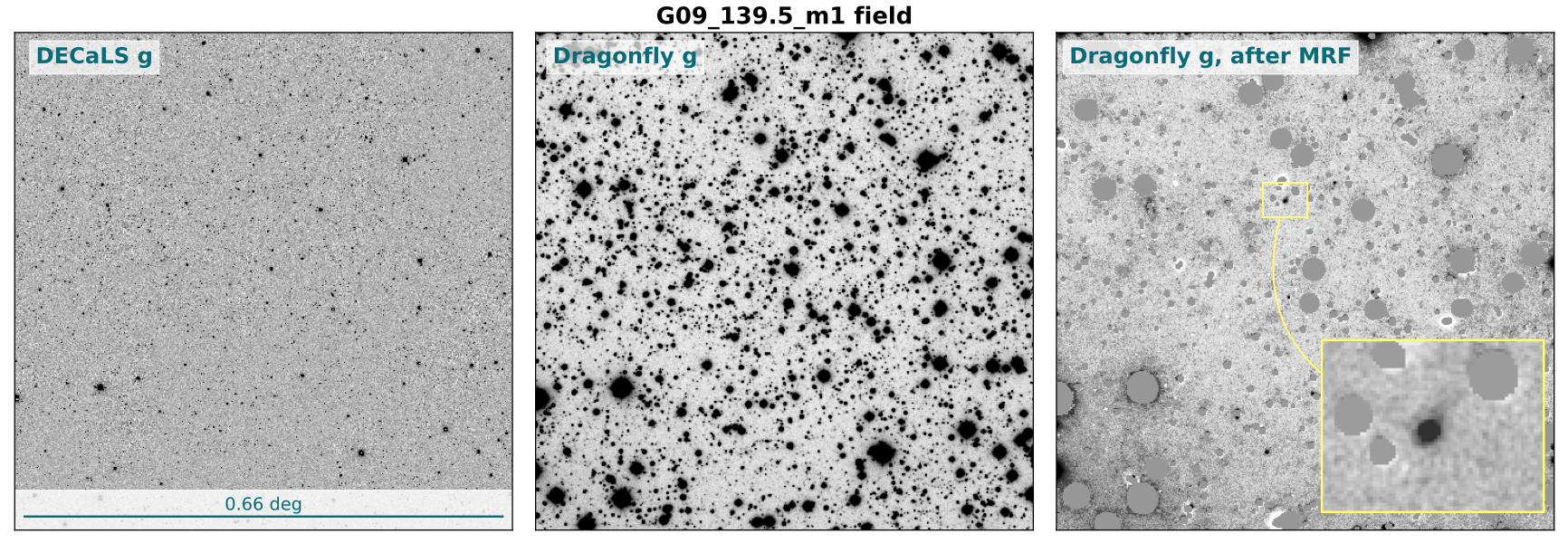}
  \caption{A $0.66 \times 0.66 \ \mathrm{deg}^2$ cutout of the survey field, G09\_139.5\_m1 (middle). The DECaLS data used in the MRF are shown in the left panel, and the Dragonfly data after applying the MRF are shown in the right panel. The inset shows a galaxy identified in the image with a noticeable tidal tail.}
  \label{fig:g09}
}
\end{figure*}

The survey footprint of $330 \ \mathrm{deg}^2$ is covered by 39 overlapping Dragonfly fields, each one with a field of view of $\sim 12$\,deg$^2$ after dithering. In Paper II in this series we will present the entire survey, as well as a catalog of low surface brightness objects. Here we present a preliminary analysis of several fields to illustrate the characteristics and quality of the data.

\subsection{Sample fields}

In Figures \ref{fig:g12}, \ref{fig:g09}, \ref{fig:cirrus} we show three example fields, representing various areas of the survey: Stripe82\_m35, G12\_175.5\_m1, and G09\_136.5\_1. The data reduction was performed with the Dragonfly pipeline, as described in detail in \S\,\ref{sec:pipeline}. The typical fraction of frames that was rejected by the pipeline is $40\%-60\%$ with the majority lost in the minimum number of detected objects required step. Key information, such as the photometric zero-point, number of stacked frames, average air mass from all frames, pixel scale, and the background value, is contained in the headers of the files. For the MRF we used the publicly available DECaLS data (\citealt{2016AAS...22831701B}; \citealt{2019AJ....157..168D}).

The chosen example fields differ in exposure time and sample a different area of the survey footprint. Figure \ref{fig:cirrus} shows an example for a survey field in the Stripe 82 footprint, Stripe82\_m35. It is the shallowest of the three, with a total equivalent exposure time of $5.3$ hr with the full 48-lens array. The number of stacked 10-minute exposures in the $r$ and $g$ bands is 849 and 679, respectively. The middle panel shows the $g$-band Dragonfly image, the left panel shows the DECaLS image used in the MRF, and the right panel shows the Dragonfly image after MRF was applied. As can be seen, the field is dominated by galactic cirrus emission. This example demonstrated the survey's ability to map low surface brightness phenomena on large scales: in the DECaLS image, the cirrus emission is entirely removed owing to the sky subtraction algorithm that was applied.

Figure \ref{fig:g12} shows a $1 \times 1 \ \mathrm{deg}^2$ cutout of a survey field in the equatorial GAMA region, G12\_175.5\_m1. The total exposure time is 10.5 hr (equivalent) with the full array; the total number of frames was 1686 and 1335 in the $r$ and $g$ bands, respectively. The three panels are as in Figure \ref{fig:cirrus}. In the right panel, we show the Dragonfly image after MRF where objects with surface brightness fainter than $25 \ \mathrm{mag \ arcsec}^{-2}$ and larger than $25$ pixels ($62\farcs5$) were kept. This example represents a region that is cirrus free down to our detection limit (representing the vast majority of the survey area), making it well suited for the detection of low surface brightness galaxies. An example is shown in the inset. 

Figure \ref{fig:g09} shows a $0.66 \times 0.66 \ \mathrm{deg}^2$ cutout of another survey field in the equatorial GAMA region, G09\_139.5\_m1, with a total exposure time of 9.1 hrs (the number of stacked 10-minute exposures in the $r$ and $g$ bands is 1439 and 1183, respectively). The three panels are as in Figures \ref{fig:cirrus} and \ref{fig:g12}. In the right panel, we show an example for a tidal tail associated with one of the identified galaxies.

\subsection{Achieved Depth}

The primary goal of the DWFS is to produce a high-quality imaging data set that is well suited for low surface brightness studies. The large field of view delivered by Dragonfly ($2 \fdg 6 \times 1 \fdg 9$) is traded off against relatively large pixels ($2\farcs5$) and low angular resolution. While the point-source depth is equivalent to that of a 1\,m telescope in $5\arcsec$ seeing, the extended emission depth is excellent. We measure the depth of the image by measuring the surface brightness contrast as described in van Dokkum et al. (\citeyear{2019arXiv191012867V}). Briefly, this method defines the surface brightness limit as the {\em contrast} on a particular spatial scale. It enables a repeatable and numerically stable way to compare surface brightness limits of different surveys and obtained with different telescopes. The method is implemented in the public Python code \texttt{sbcontrast}, part of the MRF package \footnote{\url{https://github.com/AstroJacobLi/mrf}}.

\begin{figure*}[t!]
{\centering
  \includegraphics[width=0.95\textwidth]{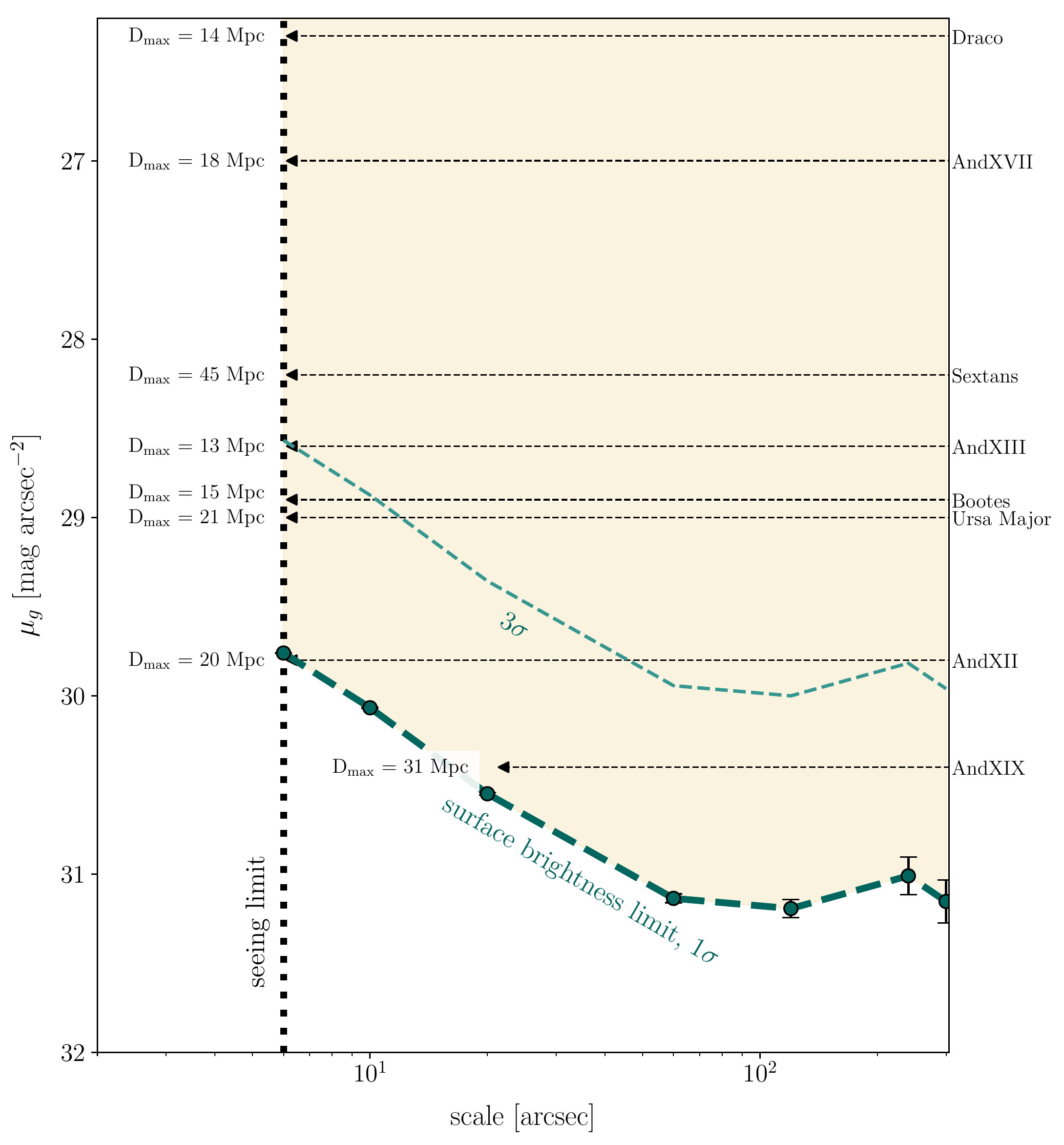}
  \caption{Measured limiting surface brightness in the Dragonfly Wide Field Survey field G12\_175.5\_m1, as a function of scale, for the $g$ band (green circles and dashed line). The seeing limit of $6''$ is shown with the vertical dotted black line. The horizontal arrows show the detectability of several Local Group ultra faint dwarf galaxies (all fainter than $26 \ \mathrm{mag \ arcsec}^{-2}$), with their maximal detectable distances based on their sizes and their mean $g$-band surface brightness within the effective radius. The G12\_175.5\_m1 field used in this calculation represents a typical survey field, with median depth comparable to other survey fields.}
  \label{fig:lim_sb}
}
\end{figure*}

For a typical survey field, the $1\sigma$ surface brightness reaches $\approx 31$\,mag\,arcsec$^{-2}$ on scales $\gtrsim 60\arcsec$, or $>3\sigma$ detection limits of galaxies with a spatial extent at least as large as this and an average surface brightness $\lesssim 29.8$\,mag\,arcsec$^{-2}$. This fulfills the science requirement of the survey (see \S\,4.3). We express the detection limit in an ability to detect analogs to particular Local Group galaxies in Figure \ref{fig:lim_sb}. The horizontal lines show the average surface brightness within the effective radius for these galaxies, with the maximal detectable distances, taking into account the seeing limit and the $1\sigma$ surface brightness limit. The dwarf galaxies Bootes, Ursa Major, Sextans, and Draco should be readily detectable out to $10-20$\,Mpc.

\subsection{Examples of Low Surface Brightness Objects}

\begin{figure*}[t!]
{\centering
  \includegraphics[width=\textwidth]{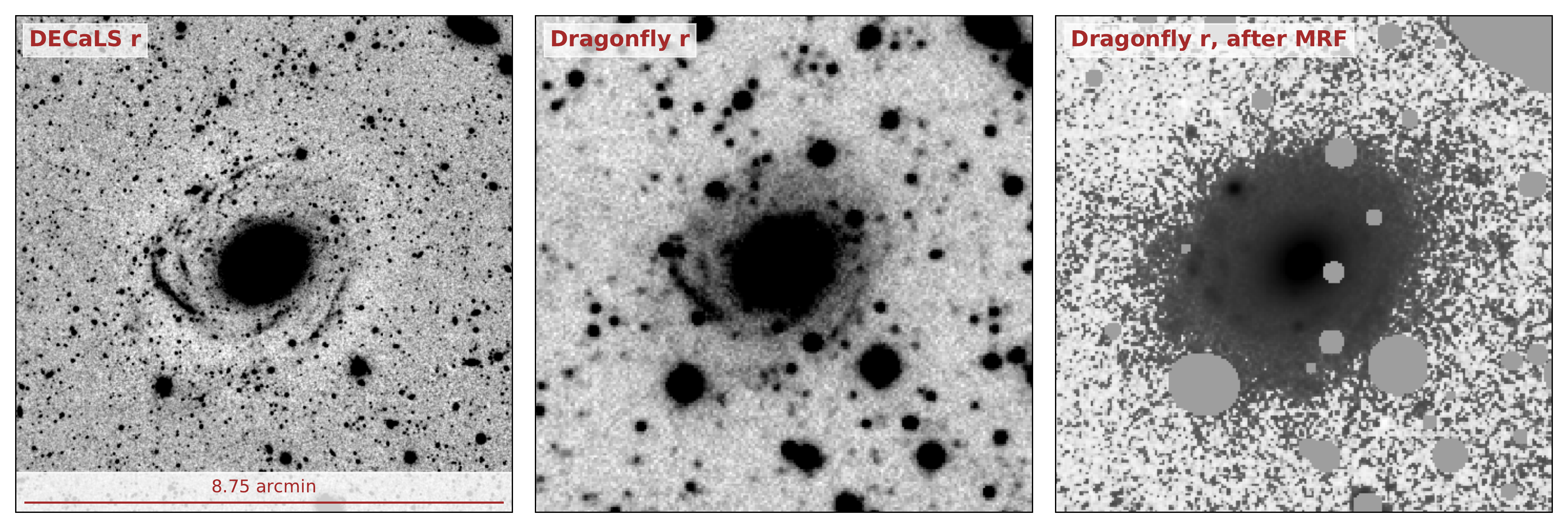}
  \caption{Giant low surface brightness galaxy UGC 1382 at a distance of $80 \ \mathrm{Mpc}$ (\citealt{2016ApJ...826..210H}) as seen in the Dragonfly Wide Field Survey (middle and right panels) and in the DECaLS data (left panel). The low surface brightness disk is nicely conserved in the Dragonfly data, where much of the light is subtracted in the DECaLS image. The right panel show the Dragonfly image after performing MRF.}
  \label{fig:example-spiral}
}
\end{figure*}

\begin{figure*}[t!]
{\centering
    \includegraphics[width=\textwidth]{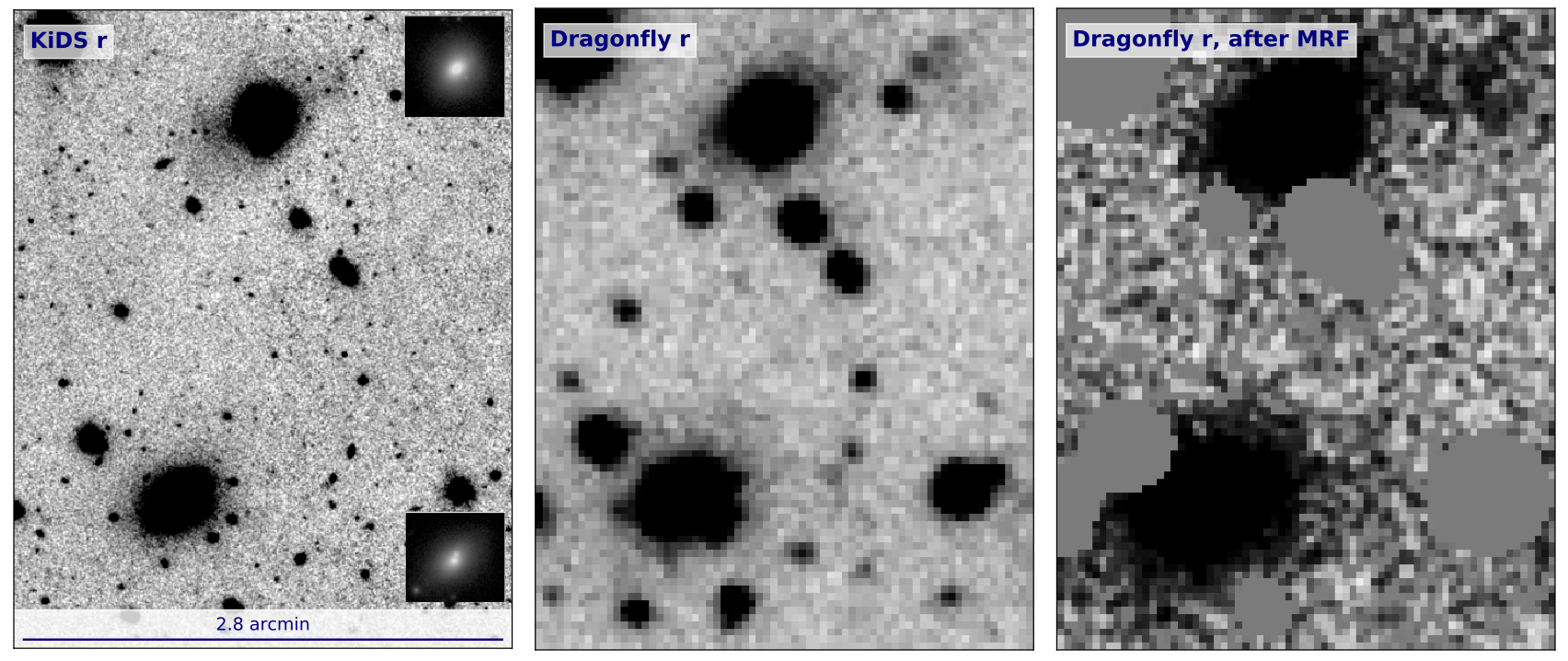}
  \caption{Two massive galaxies as seen in one of the survey fields (center). On the left panel we show the KiDS data used in the MRF process and on the right panel we show the Dragonfly data after the MRF was applied. The outskirts of the two galaxies are well detected out to larger radii than in the KiDS data. This complimentary information is useful for more accurate photometry and sizes of bright galaxies.}
  \label{fig:two-massive-galaxies}
}
\end{figure*}

In Figures \ref{fig:example-spiral} and \ref{fig:two-massive-galaxies} we present examples of low surface brightness emission detected in various survey fields.
The first is the giant low surface brightness spiral galaxy, UGC 1382, which falls in our survey area. This galaxy was long thought to be an early-type system, but in 2016 a giant disk was discovered in GALEX data (\citealt{2016ApJ...826..210H}). The left panel of Figure \ref{fig:example-spiral} shows the DECaLS image used in the MRF process, and the middle and right panels show the Dragonfly data. We detect smooth light in between the spiral arms in the galaxy and a larger extent of the disk than was previously known.

Figure \ref{fig:two-massive-galaxies} highlights the role of Dragonfly for detecting the low surface brightness outskirts of galaxies. The two galaxies in this image are embedded in extensive and irregular stellar envelopes, which are oversubtracted in the KiDS data shown in the left panel. The KiDS and Dragonfly data are complementary: in KiDS, the inner shapes of the galaxies can be measured, and the double nucleus of the bottom object is resolved.

\section{Summary} \label{sec:summary}

In this paper we introduced the DWFS, an imaging survey carried out with the 48-lens Dragonfly Telephoto Array. DWFS has surveyed 330 $\mathrm{deg}^2$ of the well-studied Stripe 82 and GAMA fields in $g$ and $r$ bands. It was designed to overlap with numerous other imaging and spectroscopic surveys, as it provides a complementary view of those fields optimizing for the low surface brightness emission. With $5-10$ hr integration time on each field, the survey provides excellent extended emission depth. The present paper is the first in a series; it describes the telescope, data reduction pipeline, and survey design and gives a very preliminary characterization of the data.

The pipeline is described in considerable detail, as it is crucial to retain low surface brightness emission on large scales. We have developed a suite of custom Python-based data reduction tools that focus on dealing with systematic errors that often limit low surface brightness observations, such as careful sky modeling and subtraction. The pipeline was integrated into cloud services provided and supported by the CANFAR. The survey's data reduction has been done entirely in the cloud, and some parts of the data processing are run in a batch mode, allowing for handling massive amounts of data in a parallel and faster way. The primary products of the pipeline are calibrated, flat-fielded co-adds in $g$ and $r$. 

The DWFS was designed and optimized for mapping the population of dwarf galaxies beyond the Local Group ($\lesssim 10 \ \mathrm{Mpc}$) down to a surface brightness of $\sim 29.5 \ \mathrm{mag \ arcsec}^{-2}$ on scales of $>5''$ (\citealt{2018ApJ...856...69D}). The survey data should also benefit many other scientific topics, including the identification of the nearest UDGs, the study of stellar halos, the faint outskirts of massive galaxies, IGL, ICL, etc. A new method, MRF (\citealt{2019arXiv191012867V}), was applied to the data, efficiently isolating the faint, extended emission. 

In this paper, we demonstrate the data quality in several of the survey fields. A full description of the survey data, including a catalog of all low surface brightness objects in the survey area, will be presented in Paper II. All data, including a catalog of the low surface brightness galaxies, will eventually be made publicly available via the survey website: dragonfly-wide.com.

\acknowledgments
\section*{Acknowledgments}

The authors thank the excellent and dedicated staff at the New Mexico Skies Observatory. We also thank S\'{e}bastien Fabbro for his support and help with the CANFAR services. Support from NSF grant AST1613582 is gratefully acknowledged. 

\software{SExtractor (\citealt{1996A&AS..117..393B}), SCAMP (\citealt{2006ASPC..351..112B}), SWARP (\citealt{2002ASPC..281..228B}), \texttt{mrf} (\url{https://github.com/AstroJacobLi/mrf}), \texttt{astropy} (\citealt{2013A&A...558A..33A}), \texttt{matplotlib} (\citealt{2007CSE.....9...90H}), \texttt{numpy} (\citealt{2011CSE....13b..22V})}.

\end{document}